\documentclass[runningheads]{llncs}
\usepackage[T1]{fontenc}

\usepackage{amsmath}
\usepackage{utfsym}

\usepackage{url}

\usepackage{breakurl}
\usepackage[breaklinks]{hyperref}
\usepackage{hyperref}
\usepackage{graphicx}
\usepackage{fontawesome}

\makeatletter
\newcommand{\printfnsymbol}[1]{%
  \textsuperscript{\@fnsymbol{#1}}%
}
\makeatother
\begin{document}

%
% \title{Contribution Title\thanks{Supported by organization x.}}
\title{GARDNet: Robust Multi-View Network for Glaucoma Classification in Color Fundus Images}
\titlerunning{Multi-View Network for Glaucoma Classification}
% If the paper title is too long for the running head, you can set
% an abbreviated paper title here
%
\author{Ahmed Al Mahrooqi\printfnsymbol{1} \and Dmitrii Medvedev\thanks{Equal contribution.} \and
Rand Muhtaseb\printfnsymbol{1}  \and Mohammad Yaqub}
%index{Al Mahrooqi, Ahmed}
%index{Medvedev, Dmitrii}
%index{Muhtaseb, Rand}
%index{Yaqub, Mohammad}

\authorrunning{A. Mahrooqi et al.}

\institute{
Mohamed bin Zayed University of Artificial Intelligence\\ Abu Dhabi, UAE\\
\email{\{ahmed.mahrooqi, mohammad.yaqub\}@mbzuai.ac.ae}}

\maketitle              % typeset the header of the 
\begin{abstract}

Glaucoma is one of the most severe eye diseases, characterized by rapid progression and leading to irreversible blindness. It is often the case that diagnostics is carried out when one's sight has already significantly degraded due to the lack of noticeable symptoms at early stage of the disease. Regular glaucoma screenings of the population shall improve early-stage detection, however the desirable frequency of etymological checkups is often not feasible due to the excessive load imposed by manual diagnostics on limited number of specialists. Considering the basic methodology to detect glaucoma is to analyze fundus images for the \textit{optic-disc-to-optic-cup ratio}, Machine Learning algorithms can offer sophisticated methods for image processing and classification. In our work, we propose an advanced image pre-processing technique combined with a multi-view network of deep classification models to categorize glaucoma. Our \textit{Glaucoma Automated Retinal Detection Network (GARDNet)} has been successfully tested on Rotterdam EyePACS AIROGS dataset with an AUC of 0.92, and then additionally fine-tuned and tested on RIM-ONE DL dataset with an AUC of 0.9308 outperforming the state-of-the-art of 0.9272. Our code is available on \url{https://github.com/ahmed1996said/gardnet}

\keywords{Glaucoma Classification  \and Color Fundus Images \and Computer Aided Diagnosis \and Deep Learning.}
\end{abstract}

\section{Introduction}
Glaucoma is an eye disease which is considered the leading cause of blindness. It is caused by an increased pressure in the eyes as a result of fluid build up, clinically known as \textit{intraocular pressure (IOP)}, which damages the optic nerve. Patients with glaucoma do not usually experience symptoms, as such, it is referred to as the “silent thief of sight”  \cite{Lee2005}. A recent study \cite{Allison} reported that by the year 2040, 111.8 million people will be affected by this disease. Among many types of glaucoma, there are two common types, specified by the structural nature of the disease: \textit{angle closure glaucoma (ACG)} and \textit{open angle glaucoma (OAG)}. The former type is more common, while the latter progresses much faster to complete blindness with no early intervention. While measuring the IOP may sometimes help clinicians in diagnosis, it is difficult to take accurate readings due to the unstable nature of the optical pressure. Clinicians have resorted to examining the structure and appearance of \textit{optic disc (OD)}, such as the increase of the \textit{cup-to-disc ratio (CDR)} \cite{Lee2005}:  the ratio of the optic cup diameter to the diameter of the OD. However, manual examination is time-consuming and is a subject to the availability of a specialist. In order to release optometrists and ophthalmologists from the burden of manual glaucoma screening, multiple deep learning approaches are explored.

In this paper, we propose \textit{Glaucoma Automated Retinal Detection Network (GARDNet)}: a combined methodology of sophisticated image pre-processing and robust multi-view network architecture for glaucoma classification. Our model  was trained and tested on AIROGS training dataset \cite{airogs-dataset} with images of different quality and resolution. In order for our model to produce consistent and robust results regardless of the input's quality, we introduced a localization of the area of interest with the following pre-processing pipeline. GARDNet extracts bounding boxes around the OD, and then applies multiple random affine and non-linear transformations, as well as such image processing techniques as \textit{Contrast Limited Adaptive Histogram Equalization (CLAHE)}. Overall, we have validated eight models with over 150 experiments, and combined three best  performing models in a multi-view network manner. Our proposed methodology allowed us to achieve AUC of 0.9308 on an external testing dataset, outperforming the state-of-the-art model by Fumero et al. \cite{rim-one-dl} which achieved 0.9272 on the same dataset. This work does not aim to propose a new algorithm nor expand on an existing one. We aim to propose and validate a robust solution for glaucoma classification.

\section{Related Works}
Glaucoma related research is mainly focused on automated screening methods and OD segmentation as well as its outer area, with the following classification of referable/no-referable glaucoma. For example, Dibia et al. in \cite{Dibia} proposed to extract from segmented OD such features of eye fundus images as OD area, cup diameter, rim area and other important features to calculate \textit{then Cup-to-disc ratio (CDR)}, which is commonly used as glaucoma indicator. Although the proposed methodology has a strong logical foundation, it was tested on a rather small dataset. 
% and may be sensitive to the quality of fundus images collected in real life conditions.
Furthermore, many papers introduce deep-learning methods to classify glaucoma. Lee et al. in \cite{Joonseok} proposed fully automated CNN, called M-Net, based on a modified U-Net \cite{unet} to segment OD and \textit{optic cup (OC)}. For the glaucoma classification task, the team used pretrained ResNet50 and affine transformations for image preprocessing, achieving AUC of 0.96 on the small dataset, REFUGE \cite{REFUGE}.
Similar approach of two-step glaucoma screening was presented by Sreng et al. \cite{Sreng}, segmenting OD with DeepLabV3, and then classifying glaucoma with various deep CNNs such as AlexNet, GoogleNet, and InceptionV3. The authors worked with several datasets and achieved promising results, but faced some limitations when generalizing between datasets. 
%Another limitation was related to the quality of images, which in real screening setting can be lower than expected by the model. 
Maadi et al. in \cite{Maadi} followed the same segment-and-classify approach. As a novelty, the authors modified classical U-Net model, introducing pre-trained SE-ResNet50 on the encoding layers, which achieved better results.

In a more recent work, Phasuk et al. in \cite{automated-ensemble}, proposed improvements of \textit{disc-aware ensemble network}\textit{ (DENet)} which incorporate the information from general fundus image with the information from optic disc area. This allowed to achieve AUC of 0.94 on a combined testing set from RIM-ONE-R3 \cite{rim-one-r3} and Drishti-GS \cite{Drishti-GS}.

\section{Method}
\subsection{Preprocessing}

\begin{figure}[!t]
    \begin{minipage}{0.475\textwidth}
        \makebox[\textwidth]{\includegraphics[width=1.00\textwidth]{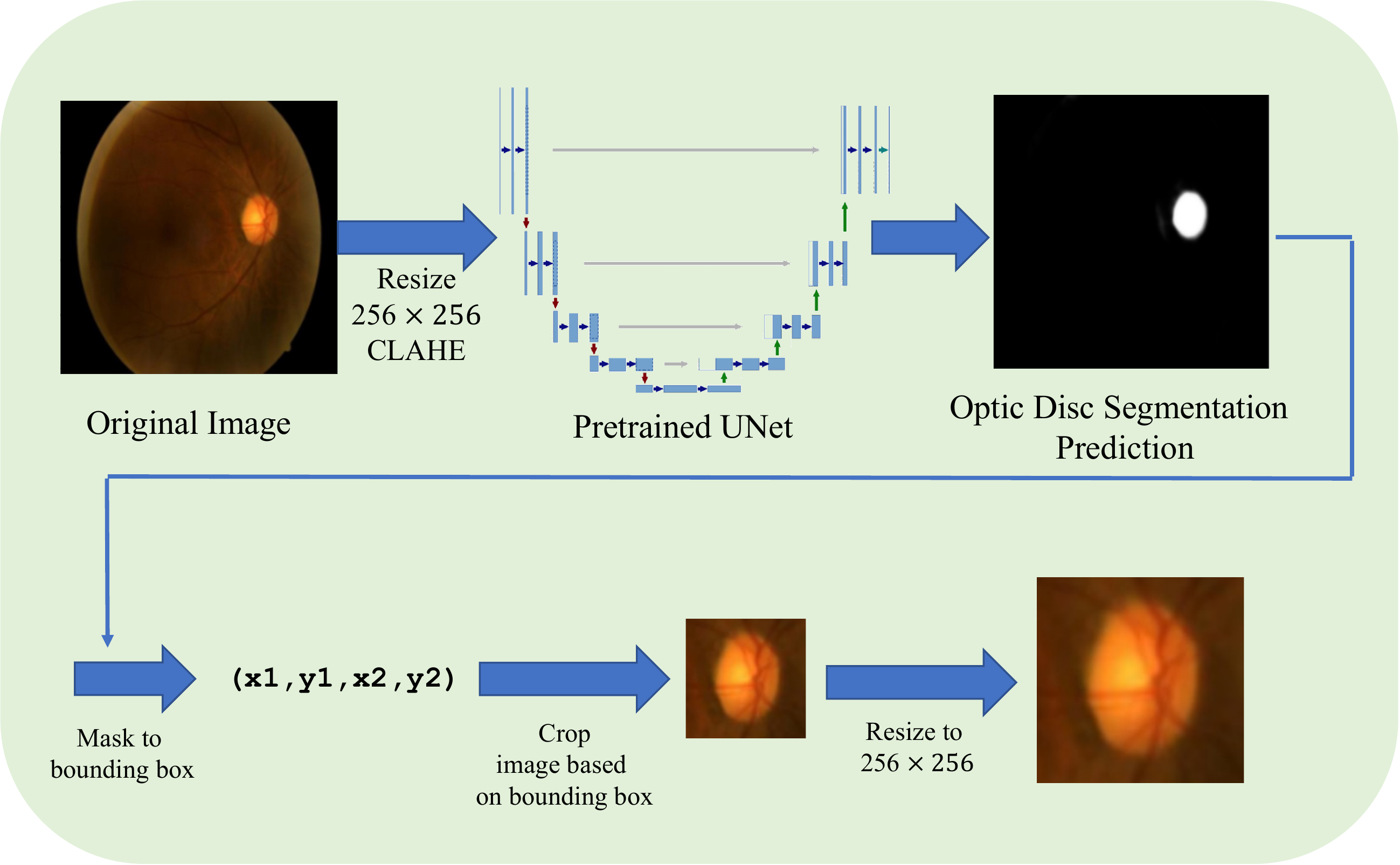}}
        \caption{Preprocessing pipeline used in our experiments on the AIROGS dataset to crop the regions of interest from the original images.}
        \label{fig:preporcessing}
    \end{minipage}%
    \hspace{0.5cm}
    \begin{minipage}{0.475\textwidth}
        \makebox[\textwidth]{\includegraphics[width=1.05\textwidth]{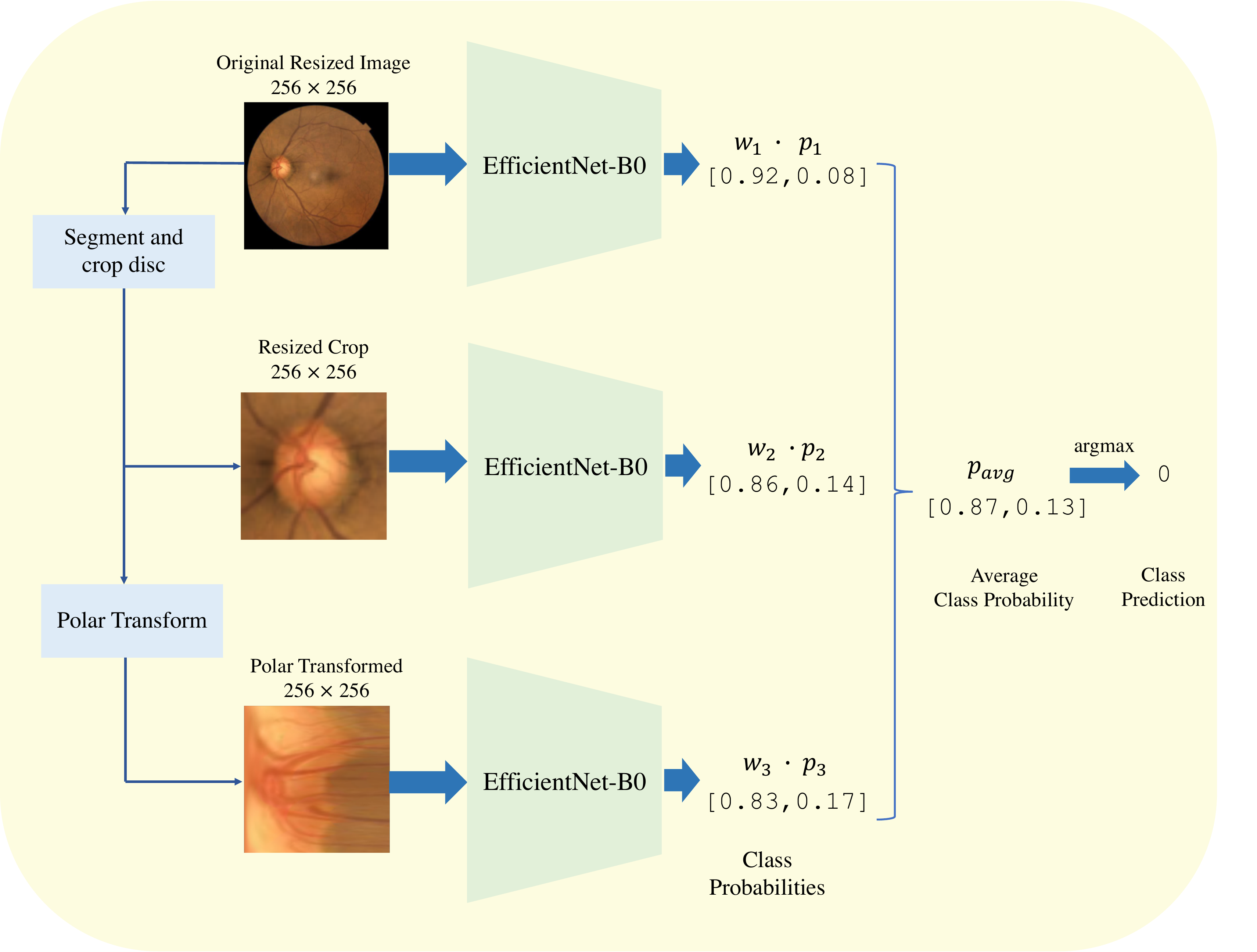}}
        \caption{Our multi-view network is composed of three different CNNs trained on different views of the color fundus images. 
        % The classification result is obtained by a weighted averaging of the class probabilities in each model. 
        $w_n$ refers to the weight coefficient corresponding to model $n$, and $p_n$ is the soft-max probability predictions in model $n$. }
        \label{fig:architecture}
    \end{minipage} 
\end{figure}

The AIROGS dataset \cite{airogs-dataset} used in our experiments have non-uniform dimensions. We therefore begin by resizing all images to a fixed dimension of 256 \(\times\) 256 pixels. In addition, we apply  CLAHE transformation with a clipping limit value of 0.01, and then feed our resized images to a U-Net model \cite{unet} pretrained on optic disc segmentation \cite{Sevastopolsky_2017} using the RIM-ONE v3 dataset \cite{rim-one-v3}. The generated optic disc segmentation masks are then converted to bounding box coordinates, with padding determined by taking 30\% of the segmented optic disc's diameter. We then proceed with cropping the original image based on the values of the bounding box coordinates, and finally resize the cropped image to a uniform dimension of 256 \(\times\) 256 pixels. In the case where the pretrained network fails to segment the optic disc, we default by taking a center crop of size 85\% of the image width, followed by a resize to 256 \(\times\) 256 pixels. This accounts for approximately 20\% of our dataset. \autoref{fig:preporcessing} illustrates our overall preprocessing pipeline.

\subsection{Multi-View Classification Network}
Our glaucoma classification model, GARDNet, is composed of a multi-view network of three different convolutional neural networks (CNNs) trained on different views of the color fundus images, as illustrated in \autoref{fig:architecture}. The first network is trained on the original resized images, whereas the second network is trained on the cropped disc area generated from the preprocessing strep, and finally, the third network is trained on the polar transformed cropped images. The training of each model is done independently. The model choice in the final multi-view network is based on ablation studies using different architectures, as reported in later sections. The intuition behind the multi-view network is that, experimentally, the model with uncropped images performed better than the cropped images. This is likely due to the error introduced by the pretrained disc segmentation model that is used to crop the images. At the same time, cropped images containing the optic disc area are most important for glaucoma diagnosis, as stated in the literature \cite{Joonseok} and shown experimentally in our GradCAM \cite{GradCAM} visualization \autoref{fig:gradcam-uncropped}. We therefore retain both models in the final multi-view network. Lastly, in the final model, we apply polar transformation, which converts the image representation from Cartesian coordinates to polar coordinates system. For a point $(u,v)$ in the Cartesian space, we obtain the radius $r$ and angle $\theta$ as follows \cite{polar}:

\begin{equation}
 \begin{cases} 
      r = \sqrt{u^2 + v^2} \\
      \theta = \tan^{-1}({\frac{v}{u}})
   \end{cases}
    \leftrightarrow
   \hspace{1.5mm}
 \begin{cases} 
      u = r \cos{\theta} \\
      v = r \sin{\theta}
   \end{cases}
\end{equation}

The transformation converts the radial relationship between the optic disc, cup, and background to a spatial hierarchical structure, which may provide an alternative view to the classification model and help capture more complex features. Phasuk et al. \cite{automated-ensemble} claims that this transformation enhances the low level information in the optic disc region. The final classification prediction is obtained by taking a weighted average of the three soft-max predictions, followed by assigning the prediction label to the class that scored the highest probability. In the final multi-view network, we assign higher weight ($w = 2$) to the model trained on uncropped images, as it performed better on the validation set. The other two models generally performed similarly and therefore share the same weight.

\begin{figure}[t]
\centering\includegraphics[width=.85\textwidth]{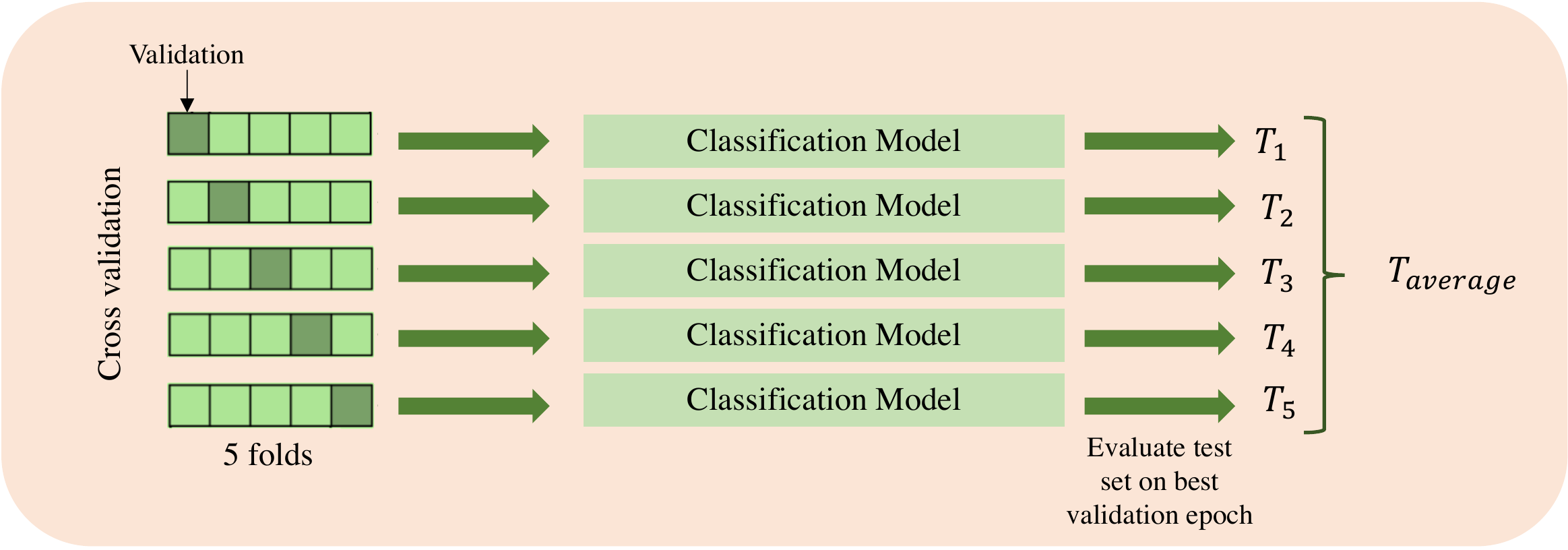}
\caption{5-fold cross-validation applied on our dataset, where \(T_n\) is the test results evaluated using the model trained on fold \(n\) and \(T_{average}\) is the average test results across all \(k=\)5 folds. } 
\label{fig:cross-val}
\end{figure}

% \section{Experiments - Rand}
% In this section, we aim to give a brief summary of the datasets used, experimental setup and all the experiments that are performed on the two datasets.
\section{Datasets}

\subsubsection{Rotterdam EyePACS AIROGS} The Rotterdam EyePACS AIROGS dataset \cite{airogs-dataset} consists of 113,893 color fundus images. Only the training set is public and available to be downloaded, which has 101,442 gradable images (images of acceptable quality). The testing set consists of 11,000 gradable and ungradable images but it is not accessible to the public which limited our ability to use in this paper. Each image in the dataset is annotated by an expert as “referable glaucoma” or “no referable glaucoma”. The images are high in resolution and do vary in size. The dataset has significant class imbalance, where the size of “no referable glaucoma” (normal) class is approximately 15 times greater than the “referable glaucoma” class.

\subsubsection{RIM-ONE DL} Retinal IMage database for Optic Nerve Evaluation for Deep Learning (RIM-ONE DL) dataset \cite{rim-one-dl} is used in this project as an external testing dataset, which consists of 313 normal and 172 glaucomatous fundus images. All images were segmented, then cropped around the cup-disc area. There are two training/testing split versions of this dataset; one was split randomly and the other was split by hospitals in Madrid and Zaragoza. We chose to report the results on the one split by the hospitals. The training set consists of 311 images, and the testing set contains 174 images.

\section{Experimental Setup}

For the following datasets, all the images were resized to 256×256. To address the problem of the imbalanced classes in both datasets, we utilize weighted cross entropy as a loss function, with the weights assigned for class $j$ being $w_j = n_{samples}/({n_{classes}\cdot n_{samples_j}})$ . In other words, the class weights are inversely proportional to their respective class probabilities. The evaluation metrics used are receiver operating characteristic area under curve (ROC AUC) and F-score (F1).

\subsubsection{Rotterdam EyePACS AIROGS}
Our models were trained for 50 epochs on a single NVIDIA A100 GPU with a batch size of 64. An Adam optimizer was used with a learning rate ranging between $1 \times 10^{-4}$ - $1 \times 10^{-3}$. For some experiments, we apply data augmentations consisting of random vertical flip ($p=0.5$), random horizontal flip ($p=0.5$) and random rotation (degrees=($-10^{\circ},+10^{\circ}$)). In some experiments, we apply CLAHE transformation, as inspired by previous works. The advantage of using CLAHE is that it enhances the contrasts and dampens any noise amplification \cite{automated-ensemble}. Given that the original testing set is not available, we split the training data into training and testing splits, using approximately 90/10 percents. To validate the robustness of our models, we performed 5-fold cross-validation by splitting the training data into training and validation, with approximately 80/20 percent, as illustrated in \autoref{fig:cross-val}. As a result, the dataset sizes for training, validation, and testing are 73,154, 18,288 and 10,000, respectively.

\subsubsection{RIM-ONE DL}
For this dataset, our main goal is to validate our model trained on the Rotterdam EyePACS AIROGS generalizability on a completely new unseen dataset. We visually noticed that the optic disc occupied a larger area in the image. We therefore retrain our best model with scaling augmentation that mimics this behavior, and results in a model invariant to images with different scales. We fine-tuned our pretrained models using dropout rate of 0.2, learning rate of \( 5 \times 10^{-3} \) and augmentations such as random horizontal and vertical flips as well as scaling. Since the RIM-ONE DL dataset is already cropped around the optic disc (OD) area, our multi-view network for this experiment consisted of only two models, while ignoring the model trained on original uncropped images. This makes our proposed solution applicable to different datasets with varying sizes and crops.

\section{Experiments \& Results}
\subsubsection{Rotterdam EyePACS AIROGS}
\autoref{table:airogs-results} shows a summary of our experiments on the Rotterdam EyePACS AIROGS dataset. On the cropped data, we performed multiple experiments using different convolutional neural networks such as EfficinetNet-B0, EfficinetNet-B1, MobileNet-V3, ResNet18, ResNet34, ResNet50 and DenseNet in addition to Vision Transformer (ViT-Base) with patch size 16. We experimented with several experimental hyperparameters such as dropout, applying CLAHE and augmentations. Our best performance on the cropped images was obtained with EfficientNet-B0 model with dropout ($p=0.5$), CLAHE and augmentations, gaining an AUC of  0.90 $\pm$ 0.01 and F1-score of 0.77 $\pm$ 0.01. To verify the performance of our cropped images vs. the original images, we trained the original images on the same model configurations as our best model, and obtained a higher AUC of 0.91 $\pm$ 0.01 and F1-score of 0.79 $\pm$ 0.01. Additionally, we further repeated the training of the best scoring model on the uncropped images using polar transformations, which obtained a slightly inferior performance of  AUC 0.85 $\pm$ 0.02 and F1-score of 0.76 $\pm$ 0.01.

\subsubsection{RIM-ONE DL}
\autoref{table:results} summarizes our testing results on the RIM-ONE DL dataset after fine-tuning our pretrained models (Experiment \#10 and \#14) on the RIM-ONE DL training split. As shown in \autoref{table:results}, our multi-view network achieves a better performance compared to \cite{rim-one-dl}, obtaining an AUC of 0.9308 and F1-Score 0f 0.9170. The individual CNNs corresponding to Experiments \#10 and \#14 obtained an inferior performance of AUC 0.87954 and 0.9088 respectively.

\begin{table}[t]
\centering
\caption{Experimental results on the Rotterdam EyePACS AIROGS dataset. \textit{D=}dropout, \textit{A=}augmentations, \textit{C=}CLAHE, \textit{S=}scaling transformation, \textit{P=}polar transformation. EfficientNet-B0$_{\text{original}}$ refers to EfficientNet-B0 model trained on the original (uncropped) data.}\label{table:airogs-results}
\begin{tabular}{l|c|c|c|c|c|c|c|c}
\hline
\textbf{ID} & \textbf{Model} & \textbf{D} & \textbf{A} & \textbf{C} & \textbf{P} & \textbf{S}  &\textbf{Test AUC}& \textbf{Test F-1}\\
\hline
% 1 & ViT-B$_{224 \times 224}$ (D,A,C) & 0.65 $\pm$ 0.05  & 0.57 $\pm$ 0.04\\
1 & Resnet34 & & & & &  & 0.74 $\pm$ 0.02 & 0.74 $\pm$ 0.01\\
2 & Resnet18 & & & & &  & 0.75 $\pm$ 0.02 & 0.74 $\pm$ 0.01\\
3 & Resnet34    & & & \usym{1F5F8} & &  & 0.78 $\pm$ 0.01 & 0.75 $\pm$ 0.01 \\
4 & ViT-B$_{224 \times 224}$ & & & & &   & 0.78 $\pm$ 0.01 & 0.75 $\pm$ 0.01\\
5 & DenseNet-121  & & & & &  & 0.79 $\pm$ 0.04 & 0.71 $\pm$ 0.06\\
6 & MobileNet-V3 Large  & & & & & & 0.79 $\pm$ 0.02 & 0.78 $\pm$ 0.00\\
7 & EfficientNet-B1  & & & & &  & 0.81 $\pm$ 0.02 & 0.80 $\pm$ 0.01\\
8 & EfficientNet-B0  & & & & &  & 0.81 $\pm$ 0.02 & 0.79 $\pm$ 0.01\\
9 & Resnet50   & & & & &  & 0.82 $\pm$ 0.02 & 0.80 $\pm$ 0.00 \\
10 & EfficientNet-B0 &\usym{1F5F8} &\usym{1F5F8} &\usym{1F5F8} &\usym{1F5F8} &  & 0.85 $\pm$ 0.02 & 0.76 $\pm$ 0.01\\
11 & EfficientNet-B0  & &\usym{1F5F8} &\usym{1F5F8} & &  & 0.87 $\pm$ 0.01 & 0.80 $\pm$ 0.01\\
12 & MobileNet-V3 Large  & \usym{1F5F8}&\usym{1F5F8} &\usym{1F5F8} & &  & 0.89 $\pm$ 0.01 & 0.77 $\pm$ 0.01\\
13 & EfficientNet-B0  & \usym{1F5F8}& \usym{1F5F8}& \usym{1F5F8}& & \usym{1F5F8} & 0.89 $\pm$ 0.01 & 0.76 $\pm$ 0.01 \\
14 & EfficientNet-B0  &\usym{1F5F8} &\usym{1F5F8} &\usym{1F5F8} & &  & 0.90 $\pm$ 0.01 & 0.77 $\pm$ 0.01\\
15 & EfficientNet-B0$_{\text{original}}$ & \usym{1F5F8}&\usym{1F5F8} &\usym{1F5F8} & &   & 0.91 $\pm$ 0.01 & 0.79 $\pm$ 0.01\\
\hline
16 & Multi-view network (\#14, \#15) & & & & &   & \textbf{0.92} & 0.80 \\
17 & Multi-view network (\#10, \#14, \#15) & & & & &   & \textbf{0.92} & 0.80 \\
\hline
\end{tabular}
\end{table} 

\begin{table}
\centering
\caption{Experimental results on RIM-One DL dataset.}\label{table:results}
\begin{tabular}{l|c|c}
\hline
\textbf{Model} &\textbf{Test AUC}& \textbf{Test F-1}\\
\hline
Fumero et al. \cite{rim-one-dl} & 0.9272 & - \\
\hline
% EfficientNet-B0 & 100 & ImageNet & 0.72 & 0.75 \\
% \hline
% EfficientNet-B0 & 0& Experiment \#13 &0.66 & 0.67\\
% EfficientNet-B0 & 25& Experiment \#13 & 0.86 & 0.87\\
% EfficientNet-B0 & 50& Experiment \#13 &0.89 & 0.89\\
Experiment \#10 (Ours)  & 0.8795 & 0.8517\\
Experiment \#14 (Ours)  & 0.9088 & 0.8974\\
Multi-view network (\#10, \#14) (Ours)  &\textbf{0.9308} & \textbf{0.9170}\\
% 207/62 & EfficientNet-B0 (D, A, C, S) & xx & xx\\
\hline
\end{tabular}
\end{table}

\section{Discussion}

Our results indicate that the models trained on the uncropped images performed much better than the cropped images. We hypothesize that this is due to errors introduced by the pretrained disc segmentation model that is used to crop the images. Furthermore, our experiments show that using dropout improves the performance, and therefore  is a good strategy for overfitting. Additionally, using augmentations and CLAHE on top of dropout significantly improves the performance. By using augmentations and CLAHE, we increase the effective dataset size and also overcome model overfitting, making it robust to spatial and color transformations.

Furthermore, the ViT-B model did not perform as good as the EfficientNet-B0 model. The Vision Transformer model requires a large amount of data to perform as well as CNNs, therefore we hypothesize that its inferior performance is probably due to the relatively small number of training samples. In addition, ResNet18 and ResNet34 models performed worse, as smaller models are not able to capture the complex features in our dataset.

The multi-view network outperformed all previous experiments in the AIROGS dataset. By combining our three best performing CNNs, each trained on a different view of the same data, we achieve an AUC of 0.92. Furthermore, we give more classification decision weight to the best performing CNN, Experiment \#15 (\autoref{table:airogs-results}), which helped achieve this performance.

For the RIM-ONE DL experiments, we can conclude that our multi-view network, GARDNet, generalizes well on this dataset when fine-tuned on the training set. While experiment \#10 and \#14 did not exceed in performance compared to the previous state-of-the-art, the multi-view network composed of these two models scored a higher AUC score than Fumero et al. \cite{rim-one-dl}. We hypothesize that our model performed better due to our image processing methods such as CLAHE and polar transformations, as well as the availability of a large dataset for pretraining. Furthermore, as our results indicate, multi-view networks outperform individual models. 

Finally, we address the results obtained in Experiment \#16 from \autoref{table:airogs-results}. As we can see, on the AIROGS dataset, our multi-view model without the polar network performs as good as the Multi-view model with all three networks (Experiment \#16). While this may indicate that the polar network has no positive contribution to the overall model, we argue that our results on the external dataset prove the opposite. As shown in \autoref{table:results}, the polar model alone (Experiment \#10) had achieved inferior performance in comparison to the cropped model (Experiment\#14), but when combined in a multi-view network, the result achieved is significantly higher than the individual networks. 

\section{Conclusion}
In this paper, we introduced a multi-view network GARDNet for glaucoma classification composed of three different CNNs trained on different views of color fundus images. Trained on the AIROGS dataset and tested on an external dataset, RIM-ONE DL, our results indicate that the multi-view network significantly improves the performance when compared to individual models. On the external test dataset, we get superior performance to the previous state-of-the-art model by Fumero et al. \cite{rim-one-dl}. In future works, we would like to extend the weighted averaging of the multi-view network predictions, such that the weights are determined systematically as learnable parameters rather than being constant.
% Additionally, we would like to evaluate the performance of GARDNet using smaller fractions of the training dataset, to understand the impact of dataset size to our model performance.

\newpage

\bibliographystyle{splncs04}
\bibliography{paper29.bib}

\begin{thebibliography}{10}
\providecommand{\url}[1]{\texttt{#1}}
\providecommand{\urlprefix}{URL }
\providecommand{\doi}[1]{https://doi.org/#1}

\bibitem{Allison}
Allison, K., Patel, D., Alabi, O.: Epidemiology of {Glaucoma}: {The} {Past},
  {Present}, and {Predictions} for the {Future}. Cureus  (Nov 2020).
  \doi{10.7759/cureus.11686},
  \url{https://www.cureus.com/articles/42672-epidemiology-of-glaucoma-the-past-present-and-predictions-for-the-future}

\bibitem{Dibia}
Dibia, A.C., Nwawudu, S.E.: Automated detection of glaucoma from retinal images
  using image processing techniques  \textbf{7},  2321--9009 (09 2018)

\bibitem{polar}
Fu, H., Cheng, J., Xu, Y., Wong, D.W.K., Liu, J., Cao, X.: Joint optic disc and
  cup segmentation based on multi-label deep network and polar transformation.
  {IEEE} Transactions on Medical Imaging  \textbf{37}(7),  1597--1605 (jul
  2018). \doi{10.1109/tmi.2018.2791488},
  \url{https://doi.org/10.1109%2Ftmi.2018.2791488}

\bibitem{REFUGE}
Fu, H., Li, F., Orlando, J.I., Bogunović, H., Sun, X., Liao, J., Xu, Y.,
  Zhang, S., Zhang, X.: Refuge: Retinal fundus glaucoma challenge (2019).
  \doi{10.21227/tz6e-r977}, \url{https://dx.doi.org/10.21227/tz6e-r977}

\bibitem{rim-one-dl}
Fumero, F., Diaz-Aleman, T., Sigut, J., Alayón, S., Arnay, R., Angel-Pereira,
  D.: Rim-one dl: A unified retinal image database for assessing glaucoma using
  deep learning. Image Analysis and Stereology  \textbf{39} (10 2020).
  \doi{10.5566/ias.2346}

\bibitem{rim-one-v3}
Fumero, F., Sigut, J., Alayón, Silvia andGonzález-Hernández, M., González
  de~la Rosa, M.: Interactive tool and database for optic disc and cup
  segmentation of stereo and monocular retinal fundus images (06 2015)

\bibitem{rim-one-r3}
Fumero, F., Sigut, J., Alayón, Silvia andGonzález-Hernández, M., González
  de~la Rosa, M.: Interactive tool and database for optic disc and
  cupsegmentation of stereo and monocular retinal fundus images (06 2015)

\bibitem{Lee2005}
Lee, D.A., Higginbotham, E.J.: Glaucoma and its treatment: {A} review. American
  Journal of Health-System Pharmacy  \textbf{62}(7),  691--699 (Apr 2005).
  \doi{10.1093/ajhp/62.7.691},
  \url{https://academic.oup.com/ajhp/article/62/7/691/5134357}

\bibitem{Joonseok}
Lee, J., Lee, J., Song, H., Lee, C.: Development of an end-to-end deep learning
  system for glaucoma screening using color fundus images. JAMA Ophthalmol.
  \textbf{137},  1353–1360 (2019)

\bibitem{Maadi}
Maadi, F., Faraji, N., Bibalan, M.H.: A {Robust} {Glaucoma} {Screening}
  {Method} for {Fundus} {Images} {Using} {Deep} {Learning} {Technique}. In:
  2020 27th {National} and 5th {International} {Iranian} {Conference} on
  {Biomedical} {Engineering} ({ICBME}). pp. 289--293. IEEE, Tehran, Iran (Nov
  2020). \doi{10.1109/ICBME51989.2020.9319434},
  \url{https://ieeexplore.ieee.org/document/9319434/}

\bibitem{automated-ensemble}
Phasuk, S., Poopresert, P., Yaemsuk, A., Suvannachart, P., Itthipanichpong, R.,
  Chansangpetch, S., Manassakorn, A., Tantisevi, V., Rojanapongpun, P.,
  Tantibundhit, C.: Automated glaucoma screening from retinal fundus image
  using deep learning. In: 2019 41st Annual International Conference of the
  IEEE Engineering in Medicine and Biology Society (EMBC). pp. 904--907 (2019).
  \doi{10.1109/EMBC.2019.8857136}

\bibitem{unet}
Ronneberger, O., Fischer, P., Brox, T.: U-net: Convolutional networks for
  biomedical image segmentation (2015). \doi{10.48550/ARXIV.1505.04597},
  \url{https://arxiv.org/abs/1505.04597}

\bibitem{GradCAM}
Selvaraju, R.R., Das, A., Vedantam, R., Cogswell, M., Parikh, D., Batra, D.:
  Grad-cam: Why did you say that? visual explanations from deep networks via
  gradient-based localization. CoRR  \textbf{abs/1610.02391} (2016),
  \url{http://arxiv.org/abs/1610.02391}

\bibitem{Sevastopolsky_2017}
Sevastopolsky, A.: Optic disc and cup segmentation methods for glaucoma
  detection with modification of u-net convolutional neural network. Pattern
  Recognition and Image Analysis  \textbf{27}(3),  618--624 (jul 2017).
  \doi{10.1134/s1054661817030269},
  \url{https://doi.org/10.1134%2Fs1054661817030269}

\bibitem{Drishti-GS}
Sivaswamy, J., Krishnadas, S.R., Datt~Joshi, G., Jain, M., Syed~Tabish, A.U.:
  Drishti-gs: Retinal image dataset for optic nerve head(onh) segmentation. In:
  2014 IEEE 11th International Symposium on Biomedical Imaging (ISBI). pp.
  53--56 (2014). \doi{10.1109/ISBI.2014.6867807}

\bibitem{Sreng}
Sreng, S., Maneerat, N., Hamamoto, K., Win, K.Y.: Deep {Learning} for {Optic}
  {Disc} {Segmentation} and {Glaucoma} {Diagnosis} on {Retinal} {Images}.
  Applied Sciences  \textbf{10}(14), ~4916 (Jul 2020).
  \doi{10.3390/app10144916}, \url{https://www.mdpi.com/2076-3417/10/14/4916}

\bibitem{airogs-dataset}
Coen~de Vente, Koenraad A.~Verrmeer, N.J.: Airogs: Artificial intelligence for
  robust glaucoma screening challenge. In: IEEE International Symposium on
  Biomedical Imaging. Ieee (2022)

\end{thebibliography}

\newpage
\setcounter{section}{0}
\renewcommand{\thesection}{\Alph{section}}

\appendix
\section{Supplementary Material}
\setcounter{figure}{0} 

\renewcommand{\thefigure}{\Alph{figure}.1}

\begin{figure}
    \centering
    \includegraphics[width=1.0\textwidth]{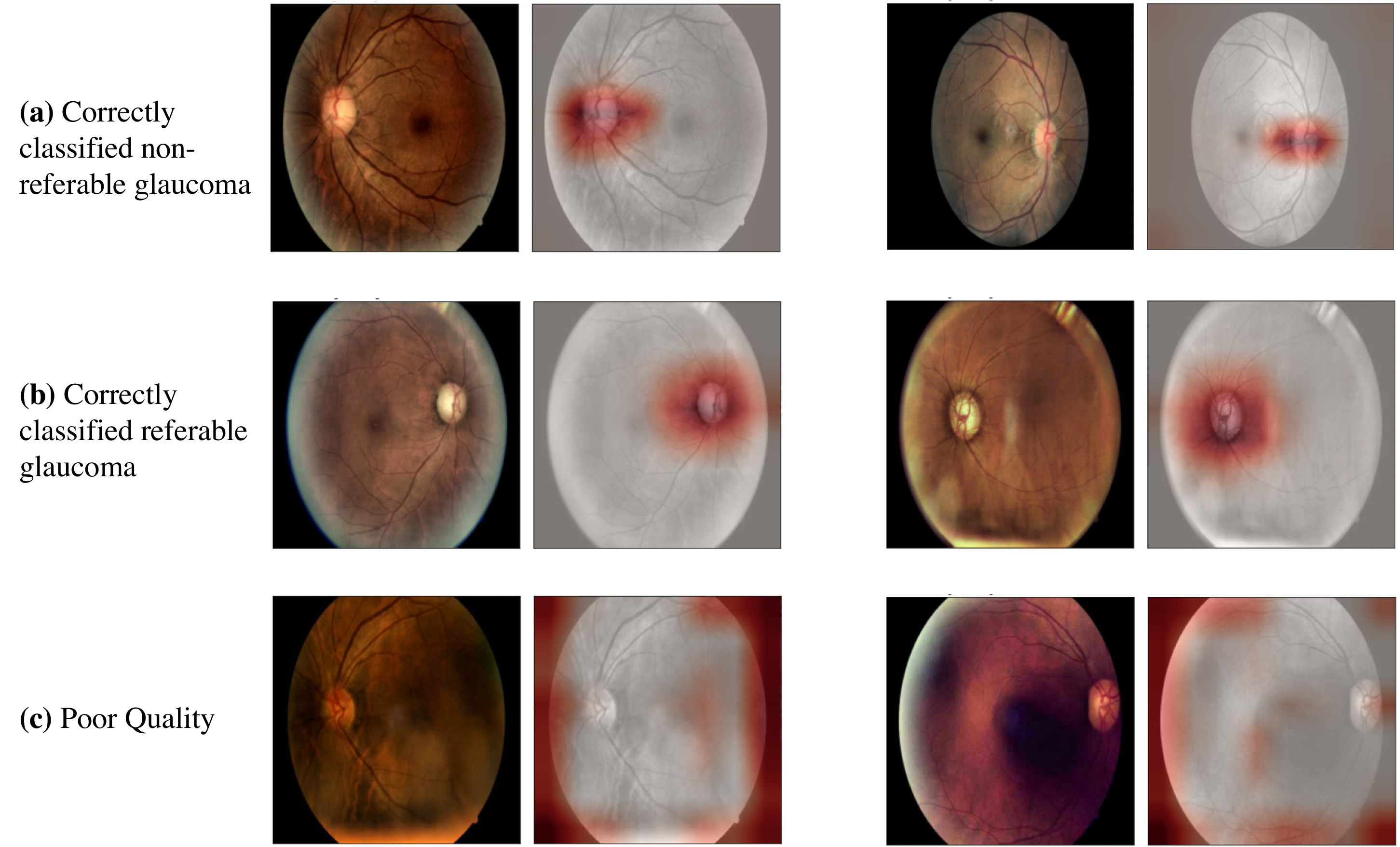}
     \caption{GradCAM \cite{GradCAM} visualization on EfficientNet-B0 model trained on the uncropped AIROGS dataset.}
    \label{fig:gradcam-uncropped}
\end{figure}
\end{document}